\documentclass[english,journal,draftcls,onecolumn,12pt]{IEEEtran}
\usepackage[T1]{fontenc}
\usepackage{babel,amsmath,amssymb,mathtools}
\usepackage{graphicx}
\usepackage{comment}
\usepackage{amsthm}
\usepackage{amsmath,amssymb,latexsym,cite}
\usepackage{pstricks, float}
\usepackage{pdfpages}
\usepackage{babel}
\usepackage{graphicx}
\usepackage{setspace}
\usepackage{footmisc}
\usepackage{algorithmic}
\usepackage{algorithm2e}
\usepackage{color, arydshln}

\newtheorem{definition}{Definition}
\newtheorem{theorem}{Theorem}

\newtheorem{lemma}{Lemma}

\renewcommand{\baselinestretch}{1}
\usepackage{algorithmic}
\usepackage{algorithm2e}
\usepackage{multicol,pdfpages,enumerate}

\usepackage{color}
\DeclareSymbolFont{lettersA}{U}{txmia}{m}{it}
 \DeclareMathSymbol{\bbr}{\mathord}{lettersA}{"92}
 \DeclareMathSymbol{\bbc}{\mathord}{lettersA}{"83}
 \DeclareMathSymbol{\bbn}{\mathord}{lettersA}{"8E}
 \DeclareMathSymbol{\bbg}{\mathord}{lettersA}{"87}

\DeclareMathAlphabet{\mathscr}{OT1}{pzc}{m}{it}
\newcommand{\proofof}[1]{\phantom{aaaaaaaa}{\em Proof of #1:}}
\def\QEDmark{\blacksquare}
\def\endproof{\hfill\QEDmark}
\newcommand{\fr}{{\textsf{\textsc{frantic}{ }}}}
\newcommand{\shofa}{{\textsf{\textsc{sho-fa-int}{ }}}}

\newcommand{\graph}{\mathcal{G}}

\newcommand{\bA}{{{\mathbf{A}}}}

\newcommand{\samp}{{ n}}
\newcommand{\spar}{{ k}}

\newcommand{\meas}{{\mu}}
\newcommand{\nm}{{m}}
\newcommand{\cnst}{{ c}}
\newcommand{\eL}{{R}}
\newcommand{\el}{{r}}
\newcommand{\set}[1]{\left\{#1\right\}}

\newcommand{\cN}{{\mathcal{N}}}

\newcommand{\cO}{{\mathcal{O}}}

\newcommand{\bd}{{\pmb{d}}}
\newcommand{\ba}{{\pmb{a}}}
\newcommand{\hbd}{{\hat{\pmb{d}}}}
\newcommand{\by}{{\pmb{y}}}

\newcommand{\cE}{{\mathcal{E}}}
\newcommand{\cP}{{\pmb{P}}}
\newcommand{\cQ}{{\pmb{Q}}}

\newcommand{\cS}{{\mathcal{S}}}

\newcommand{\cT}{{\mathcal{T}}}
\newcommand{\cV}{{\mathcal{V}}}

\newcommand{\cD}{{\mathcal{D}}}
\newcommand{\iter}{{t}}

\newcommand{\xj}{{j}}
\newcommand{\yi}{{i}}

\newcommand{\sh}[1]{\textcolor{purple}{\textbf{SC:} #1}}

\usepackage{sidecap}
\newcommand{\hide}[1]{}

\sidecaptionvpos{figure}{c}
\makeatletter
\ifCLASSOPTIONcompsoc
\usepackage[caption=false,font=normalsize,labelfont=sf,textfont=sf]{subfig}
\else
\usepackage[caption=false,font=footnotesize]{subfig}
\fi
\makeatother

\begin{document}
\title{FRANTIC: A Fast Reference-based Algorithm for Network Tomography via Compressive Sensing}
\author{{Sheng Cai \qquad Mayank Bakshi \qquad Sidharth Jaggi \qquad  Minghua Chen}\\ The Chinese University of Hong Kong
}
\maketitle
\begin{abstract}We study the problem of link and node delay estimation in undirected networks when at most $\spar$ out of $\samp$ links or nodes in the network are congested. Our approach relies on end-to-end measurements of path delays across pre-specified paths in the network. We present a class of algorithms that we call \fr\footnote{\fr stands for {\bf F}ast {\bf{R}}eference-based {\bf{A}}lgorithm for {\bf{N}}etwork {\bf{T}}omography v{\bf{I}}a {\bf{C}}ompressive sensing.}. The \fr algorithms are motivated by compressive sensing; however, unlike traditional compressive sensing, the measurement design here is constrained by the network topology and the matrix entries are constrained to be positive integers. A key component of our design is a new compressive sensing algorithm \shofa that is related to the {\textsf{\textsc{sho-fa}}} algorithm~\cite{BakJCC:12} for compressive sensing, but unlike {\textsf{\textsc{sho-fa}}}, the matrix entries here are drawn from the set of integers $\{0,1,\ldots,M\}$. We  show that $\cO(\spar\log{\samp}/\log M)$ measurements suffice both for \shofa and \fr. Further, we show that the computational complexity of decoding is also $\cO(\spar\log{\samp}/\log M)$ for each of these algorithms. Finally, we look at efficient constructions of the measurement operations through Steiner Trees.
\end{abstract}
\section{Introduction} \label{sec:intro}

Monitoring performance characteristics of individual links is important
for diagnosing and optimizing network performance. Making direct measurements
for each link, however, is impractical in large-scale networks because
(i) nodes inside the networks may not be available to carry out measurements
due to physical or protocol constraints, and (ii) measuring each link
\emph{separately} incurs excessive control-traffic overhead and is
not scalable.

A viable alternative approach is network tomography \cite{vardi1996network}.
It aims to infer the performance characteristics of internal links
by \emph{path measurements} between controllable nodes, where a path
measurement is function of the characteristics of the links on the
path. It does not require access to all the nodes and, more importantly,
it allows clever solutions to leverage the network structure ({\it e.g.},
topology and graph properties) to \emph{jointly} infer the performance
characteristics of multiple links via path measurements. Many existing
work have explored such insight to design excellent solutions that
are able to infer the congested links with much less measurements
than the direct link measurement approach~\cite{bu2002network,chen2007algebra,kleinberg2000detecting,nguyen2006using,zhao2009towards}. See \cite{castro2004network} for a survey.

Recently, Xu \emph{et al.} \cite{XuMT:11} further argue
that usually only a small fraction of network links, say $k$ out
of total $|\cE|$ links ($k\ll |\cE|)$, are congested ({\it i.e.}, experiencing
large congestion delay or high packet loss rate). They interpret each
path measurement as a linear combination of the delays or loss rates
of the $k$ congested links. With these understanding, the problem
of network tomography can be viewed as recovering a $k$-sparse link
vector from a set of linear measurements.

Exploiting the ``sparse congestion structure'' insight, Xu \emph{et
al.} \cite{XuMT:11} propose a compressive sensing%
\hide{\footnote{Compressive sensing is a recently proposed  paradigm for efficient
sample and recovery of sparse signals~\cite{CanRT:06 ,Don:06}. \hide{In a basic setting,
for a length-$|E|$ (to make the notation consistent) signal with only $k$ non-zero entries (at unknown
locations), compressive sensing says that the signal can be recovered
with high probability by using ${\cal O}(k\log |E|/k)$ linear measurements in
${\cal O}(k\log |E|/k)$ steps.} We will give a brief introduction later in Section
\ref{sec:shofa}.}} based scheme that can identify any $k$ congested links using ${\cal O}(T_\cN k\log |\cE|)$
path measurements over any sufficiently-connected graph. Here, each of the path measurement is a random walk on the graph, and $T_\cN$ is the mixing time of the random walk. Further,
they show that one can actually \emph{recover} the performance characteristics
of any $k$ congested links with again ${\cal O}(T_\cN k\log |\cE|)$ path measurements
by using $\ell_{1}$-minimization. Similar results are also
obtained by~\cite{firooz2010network,MengWMA:2012,CheKMS:2012}. Given all these exciting
results, a natural question is that can we do better and how?

\section{Our contribution}
\subsection{Summary}In this paper, we build upon our recently developed compressive sensing
algorithm named {\textsf{\textsc{sho-fa}}}~\cite{BakJCC:12} to design a new network tomography solution that we call \fr.
\fr achieves the following performance:
\begin{itemize}
\item\fr can identify any $\spar$ congested
links (or nodes) out of $\samp$ and recover the corresponding link (or node) performance characteristics
using ${\cal O}(\rho k\log{\samp}/\log{M})$ path measurements with a high probability. Here, $M\in\bbn$ and $\rho\in\Omega(1)\cap o(\samp/\spar)$ are design parameters. See Section~\ref{sec:parameters} for a discussion.\\
\item The \fr decoding algorithm  can recover the link (or node) performance characteristics
in ${\cal O}(\rho \spar\log{\samp}/\log M)$ steps.
\end{itemize}
As compared to the solution in \cite{XuMT:11}, our solution
improves both the number of measurements and the number of recovery
steps from ${\cal O}(T_{\cal N}k\log \samp)$ to ${\cal O}(k)$ (obtainable by setting $M=\cO(\samp)$).

\subsection{Techniques and results} The main techniques that lead to these improvements are as follows. First, in Section~\ref{sec:shofa}, we develop an efficient compressive sensing algorithm \shofa when the entries of the measurement matrix are constrained to be positive integers. Our algorithm borrows key ideas from a prior work~\cite{BakJCC:12} that studies compressive sensing in the unconstrained setting. A key technique here is to group together measurements and choose the ``weights'' of the measurement matrix as co-prime vectors. This ensures that each network link has a distinct signature in the measurement output, which allows us to decode the delay values for congested links in an iterative manner. Theorem~\ref{thm:shofa} states the performance guarantees of our algorithm. Next, we propose a design for measuring the delay on congested links in a network in Section~\ref{sec:linkdelay}. An important insight in our design is that by using local loops at individual edges, end-to-end delay measurements can be designed to assign different integer weights to delays for different edges. We start with a compressive sensing matrix given by \shofa and emulate the output of the matrix by first designing two correlated network paths, and then cancelling out the contribution of unwanted links by subtracting one from another. Theorems~\ref{thm:edge}-\ref{thm:link_steiner} state the performance guarantees of the \fr algorithms. We also note that the path lengths required for \fr can be suitably optimised by using Steiner Trees and network decomposition. Theorem~\ref{thm:link_steiner} and subsequent discussions point this out.
\subsection{Explanation of design parameters}\label{sec:parameters}
The parameter $M$ denotes the maximum number of times a test packet may travel over any edge. In many present day networks, the value of $M$ is usually fixed to be a small constant. In this setting, our algorithm requires $\cO(\spar\log{\samp})$ measurements and decoding steps. Additionally, if $M$ is allowed to increase with the network size (possibly, in future generation networks), the number of measurements and decoding complexity our algorithms may be decreased to $\cO(\spar)$.

The parameter $\rho$ is a design parameter that controls the tradeoff between the measurement path lengths and the number of measurements required. When $\rho=1$, we require $\cO(\spar\log\samp/\log M)$ measurements with path lengths $\cO(\samp D/\spar)$. On the other extreme, if $\rho$ is set to be $n/(k\omega(1))$, we require upto $o(n)$ measurements but with as little as $\omega(D)$ path length. In our exposition, we prove the correctness of our schemes for the case when $\rho=1$. The results for other values of $\rho$ follow from this analysis by pretending that the network has $\rho\spar$ congested nodes instead of $\spar$.

\begin{table*}[!htb]\label{tab:comparison}
\scriptsize{\small { }}
\begin{spacing}{1}\centering\begin{tabular}{|c|c|c|c|c|l|}
\hline
Reference & Type & \# Measurements & Decoding Complexity & Path Length & Network Topology\tabularnewline
\hline
\hline
\cite{MengWMA:2012} & Node & $R{\cal O}(k\log(|{\cal V}|/k))+R+1$ & cs  with $0-1$ matrix & -- & General graph, \tabularnewline
& & & & & $R$ is the radius of the graph\tabularnewline
\cline{2-6}
 & Node & ${\cal O}(rk\log(|{\cal V}|/k))+r$ & cs  with $0-1$ matrix & -- & If $G$ has an $r$-partition\tabularnewline
\cline{2-6}
 & Node & ${\cal O}(2k\log(|{\cal V}|/2k))+r$ & cs  with $0-1$ matrix & -- & Erdos-Renyi random graph $G(|{\cal V}|,p)$, \tabularnewline
& & & & &  with
$p=\beta\log|{\cal V}|/|{\cal V}|$ and $\beta\geq2$ \tabularnewline
\hline
\cite{XuMT:11} & Edge & ${\cal O}(T_{\cal N}k\log|{\cal E}|)$ & $l_{1}$ minimization & ${\cal O}(|E|/k)$ & $G$ is a $(D,c)$-uniform graph, \tabularnewline
& & & & & $D\geq D_{0}={\cal O}(c^{2}kT_{\cal N}^{2})$.\tabularnewline
\hline
\cite{firooz2010network} & Edge & ${\cal O}(k\log(|{\cal E}|/k))$ & $l_{1}$ minimization & -- & Network is 1-identifiable\tabularnewline
\hline
\cite{CheKMS:2012} & Node & ${\cal O}\left(c^{4}k^{2}T_{\cal N}^2\log(|{\cal V}|/d)\right)$ & Disjunct matrix & ${\cal O}(|{\cal V}|/(c^{3}kT_{\cal N}))$ & $G$ is a $(D,c)$-uniform graph. \tabularnewline
\cline{2-5}
 & Edge & ${\cal O}(c^{4}k^{2}T_{\cal N}^{2}\log(|{\cal E}|/d))$ & Disjunct matrix & ${\cal O}(|{\cal V}|D/(c^{3}kT_{\cal N}))$ & $D\geq D_{0}={\cal O}(c^{2}kT_{\cal N}^{2})$.\tabularnewline
\cline{2-5}
 & Node & ${\cal O}(c^{8}k^{3}T_{\cal N}^{4}\log(|{\cal V}|/d))$ & Disjunct matrix & unbounded (sink node) & \tabularnewline
\cline{2-5}
 & Edge & ${\cal O}(c^{9}k^{3}DT_{\cal N}^{4}\log(|{\cal E}|/d))$ & Disjunct matrix & unbounded (sink node) & \tabularnewline
\cline{2-6}
 & Node & ${\cal O}(k^{2}(\log^{3}|{\cal V}|))/(1-p)^{2}$ & Disjunct matrix & ${\cal O}(|{\cal V}|/(c^{3}kT_{\cal N}))$ & $G$ is $D$-regular expander graph or \tabularnewline
\cline{2-5}
 & Edge & ${\cal O}(k^{2}(\log^{3}|{\cal E}|))/(1-p)^{2}$ & Disjunct matrix & ${\cal O}(|{\cal V}|D/(c^{3}kT_{\cal N}))$ & Erdos-Renyi random graph, $G(|{\cal V}|,D/|{\cal V}|)$,\tabularnewline
\cline{2-5}
 & Node & ${\cal O}(k^{3}(\log^{5}|{\cal V}|))/(1-p)^{2}$ & Disjunct matrix & unbounded (sink node) & with $D\geq D_{0}=\Omega(k\log^{2}|{\cal V}|)$.\tabularnewline
\cline{2-5}
 & Edge & ${\cal O}(k^{3}D(\log^{5}|{\cal E}|))/(1-p)^{2}$ & Disjunct matrix & unbounded (sink node) & \tabularnewline
\hline
This & Node & ${\cal O}(k\log|{\cal V}|/\log M)$ & ${\cal O}(k\log|{\cal V}|/\log M)$ & ${\cal O}(D|{\cal V}|/k)$ & General Graph\tabularnewline
\cline{2-5}
paper & Edge & ${\cal O}(k\log|{\cal E}|/\log M)$ & ${\cal O}(k\log|{\cal E}|/\log M)$ & ${\cal O}(D|{\cal E}|/k)$ & $D$ is the diameter of the graph\tabularnewline
\hline
\end{tabular}\end{spacing}
\caption{}{{\begin{spacing}{1}\textnormal{{\bf Partial literature review}: \cite{MengWMA:2012} considers the node delay estimation problem where a set of nodes can be measured together in one measurement if and only if the induced subgraph is connected and each measurement is an additive sum of values at the corresponding nodes. The generated sensing matrix is a $0-1$ matrix, therefore the decoding complexity mainly depends on which binary compressive sensing algorithm is used. General graph and some special graphs are studied. The idea of a binary compressive sensing algorithm is borrowed by \cite{firooz2010network} where a single edge delay estimation problem is studied and the estimation is done using $l_{1}$ minimization. In \cite{XuMT:11}, a random-walk based approach is proposed to solve the $k$-edge delay estimation problem. $T_\cN$ is the $\frac{1}{(2c|\cV|)^{2}}$-mixing time of $\cN$. The networks with degree-bounded assumption are studied. Similar to~\cite{XuMT:11}, \cite{CheKMS:2012} uses random-walk measurements to solve both node and edge failure localization problem where group testing (non-linear version of compressive sensing) algorithm is used. The goal is to generate disjunct matrices which are suitable for group testing. The start points of measurements can be chosen within a fixed set of designated vertices,
or, chosen randomly among all vertices of the graph. The first type of construction which don't have the length bound covers the case that only a small subset of vertices are accessible as the starting points of the measurements. Separately, the problem of edge failure localization has also been studied in the optical networking literature~\cite{HarPWYC:2007,XuaSNT:2013,AhuRK:2011}. \cite{HarPWYC:2007}, which consider the single edge failure localization, has the same flavor as \cite{MengWMA:2012}. Binary-search type algorithms are proposed for some special graphs. For the general graphs, the upper bound on the number of measurements required for single edge failure localization is ${\cal O}(D(\cN)+\log^{2}(|\cV|))$ where $D(\cN)$ is the diameter of the graph.
In~\cite{XuaSNT:2013}, the problem of multi-link failure localization is considered. For small networks, tree-decomposition based method has the upper bound on the number of trials is $\min({\cal O}(D(\cN)\log |\cV|), {\cal O}(D(\cN)+\log^{2}(|\cV|)))$. For the large-scale networks, random-walk based method similarly to \cite{CheKMS:2012} is proposed. They also consider the practical constraints such as the number of failed links cannot be known beforehand. In \cite{AhuRK:2011}, the solution proposed is based on the $(k+2)$-edge-connected network for $k$ link failures localization.}\end{spacing}}
}
\end{table*}

\section{Model and problem formulation}

\noindent{\em \underline{Network and delay model}:}  Let $\cN=(\cV,\cE)$ be a undirected network with node set $\cV$ and
link set $\cE$. In this paper, we consider the reference-based tomography problem where the topology of $\cN$ is known. We assume that $\cN$ is connected. We say that a node $v\in \cV$ has delay $d_v$ if every packet that passes through $v$ is delayed by $d_v$. Similarly, a link $e\in \cE$ has delay $d_{e}$ if every packet passing through $e$ in any direction is delayed by $d_e$. We say a node or link is {\em congested} if the delay associated with it is non-zero. A congested node is called {\em isolated} if there exists one of its neighbours which is not congested. Let $\bd_\cV=(d_v:v\in\cV)$ and $\bd_\cE=(d_e:e\in\cE)$. Both $\bd_\cV$ and $\bd_\cE$ are unknown but have at most $\spar$ non-zero coordinates.\\
\noindent{\em \underline{Measurement model}:}  Each measurement is performed by sending test packets over pre-determined paths\footnote{In present day networks, this may be accomplished by employing source-based routing (c.f.~\cite{JohM:96}) for the test packets.} and measuring the end-to-end time taken for its transmission. Some nodes (resp. links) may be visited more than once in a given path.  As a result, each measurement output $y_{i}$, $i=1,2,\ldots,m$, is a weighted sum of delays involving nodes and links that lie in the measurement path, where, weight of a given node or link is the number of times it is visited by the measurement
path. In this paper, we consider two kinds of measurements -- {\em node measurements} and {\em link measurements}. In the node (resp. link) measurements, we associate each node (resp. link) with a real-valued delay and the objective is to reconstruct the node (resp. link) delay vector $\bd_\cV$ (resp. $\bd_\cE$) given the collection of measurement outputs.

%

{\noindent 1. \em \underline{Node measurements}:}
In the node measurement model, we associate each node with a real valued delay (see~\cite{MengWMA:2012}, for example). Let ${\cal S}\subseteq \cV$ denote a subset of nodes in $\cN$. Let $\cE_{{\cal S}}$ denote the subset of links with both ends in ${\cal S}$, then $\cN_{{\cal S}}=({\cal S},\cE_{{\cal S}})$ is the induced subgraph of $\cal N$. A set ${\cal S}$ of nodes can be measured together in one measurement if and only if $\cN_{{\cal S}}$ is connected.\\
{\noindent 2. \em \underline{Link measurements}:}
In link measurement setup, we associate each link with a {real valued} delay. Let ${\cal T}\subseteq \cE$ denote a subset of links in $\cN$. \hide{\sh{it seems that we don't need $\cN_{{\cal T}}$ in this paper, so we can delete the following denotion. }Let $\cV_{{\cal T}}$ denote the subset of nodes are the ends of ${\cal T}$, then $\cN_{{\cal T}}=(\cV_{{\cal T}},{\cal T})$
is the induced subgraph of $\cN$.} A set $\cal T$ of links can be measured together in one measurement if and only if there exists a path that traversed each link in $\cal T$.

For each of these models, we express the measurement output as a vector $\by\in\mathbb{R}^{\nm}$  that is related to the delay vector through a measurement matrix $\bA$ through matrix multiplication.\hide{ For example, in the node measurement model, $\bA$ is a $\nm\times|\cV|$  matrix whose $\yi$-th row corresponds to the $i$-th measurement, and $a_{\yi v}$ equals the the number of times the the $\yi$-th measurement path visits node $v$. The measurement vector $\by$ is related to the delay vector $\bd_\cV$ through the linear relationship $\by=\bA\bd_\cV$.}


\begin{table*}
\scriptsize{\small { }}
\begin{centering}
\begin{tabular}{|c|l|}
\hline
$n$ & Total number of links (or nodes) in the network\tabularnewline
\hline
$k$ & Number of congested links (or nodes) in the network\tabularnewline
\hline
$M$ & The maximum number of times a test packet may travel over any edge\tabularnewline
\hline
$D$ & The diameter of ${\cal N}$\tabularnewline
\hline
$T_{{\cal N}}$ & The mixing time of the random walk over graph ${\cal N}$\tabularnewline
\hline
$\rho$ & A design parameter that controls the tradeoff between the path lengths
and the number of the measurements \tabularnewline

\hline
${\cal N}$ & $\cN=(\cV,\cE)$, a undirected network with node set ${\cal V}$ and link set ${\cal E}$\tabularnewline
\hline
$d_{v}$ & Time taken by a test packet to pass through node $v\in{\cal V}$\tabularnewline
\hline
$\bd_{\cV}$ & Node delay vector of length $|\cV|$\tabularnewline
\hline
$d_{e}$ & Time taken by a test packet to pass through link $e\in{\cal E}$
in any direction\tabularnewline
\hline
$\bd_{\cE}$ & Link delay vector of length $|\cE|$\tabularnewline
\hline
\multicolumn{2}{c}{\ }\tabularnewline
\multicolumn{2}{c}{\ref{tab:notation}-A. \bf Network Parameters}\tabularnewline
\multicolumn{2}{c}{\ }\tabularnewline
\multicolumn{2}{c}{\ }\tabularnewline
\multicolumn{2}{c}{\ }\tabularnewline
\hline
$R$ & $\eL\in\bbn^+$ such that $M^\eL/\zeta(\eL)\geq 3\samp$ where $\zeta(\cdot)$ be the Riemann zeta function\tabularnewline
\hline
$\by$ & Measurement output of length $m=R\mu$ \tabularnewline
\hline
$\bA$ & Measurement matrix of dimension $R\mu\times n$\tabularnewline
\hline
$a_{ij}^{(r)}$ & The $\el$-th row entry in the $\xj$-th column of the $\yi$-th group of $\bA$ for $r=1,2,\ldots,R$, $i=1,2,\ldots,\mu$ and $j=1,2,\ldots,n$\tabularnewline
\hline
${\cal G}_{n,\mu}$ & A bipartite graph with left vertex set \{$1,2,\ldots,n$\} and right
vertex set \{$1,2,\ldots\mu$\}\tabularnewline
\hline
$N(S)$ & The set of right neighbours of a subset of left nodes $S$ in ${\cal G}_{n,\mu}$ \tabularnewline
\hline
$\cP$ & A path of length $T$ over the network $\cN=(\cV, \cE)$, {\em i.e.}, a sequence $(e_1,e_2,\ldots,e_{T})$ of links from $\cE$ \tabularnewline
\hline
$W(\cP,e)$ & The multiplicity of a link $e\in\cE$ given a path $\cP$, {\em i.e.}, the number of times $\cP$ visits $e$\tabularnewline
\hline
$\Delta(\cP)$  & The end-to-end delay for a path $\cP$
\tabularnewline
\hline
\multicolumn{2}{c}{\ }\tabularnewline
\multicolumn{2}{c}{\ref{tab:notation}-B. \bf Design Variables}\tabularnewline
\end{tabular}
\par\end{centering}

\caption{Table of notations\label{tab:notation}}

\end{table*}

\section{Key ideas}
\begin{figure}[ht]
\centering
\includegraphics[width=0.5\linewidth]{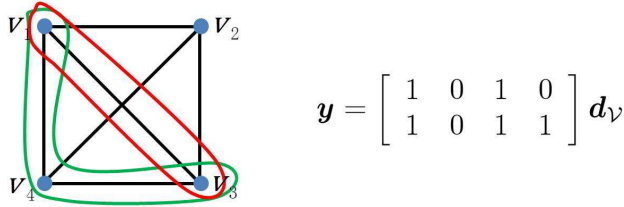}
\caption{\underline{Node Delay Estimation}: For a complete graph with four vertices. We can get any measurements we want since each subgraph of a complete graph is connected. For example, the subgraphs induced by $\{v_{1},v_{3}\}$ (covered by red cycle) and $\{v_{1},v_{3},v_{4}\}$ (covered by green cycle) are connected, therefore we get the measurements $[1\mbox{ }0\mbox{ }1\mbox{ }0]\bd_\cV$ and $[1\mbox{ }0\mbox{ }1\mbox{ }1]\bd_\cV$ respectively.}
\label{fig:nodedelay}
\end{figure}

\begin{figure}[ht]
\begin{minipage}[t]{0.48\linewidth}
\centering
\includegraphics[width=0.42\linewidth]{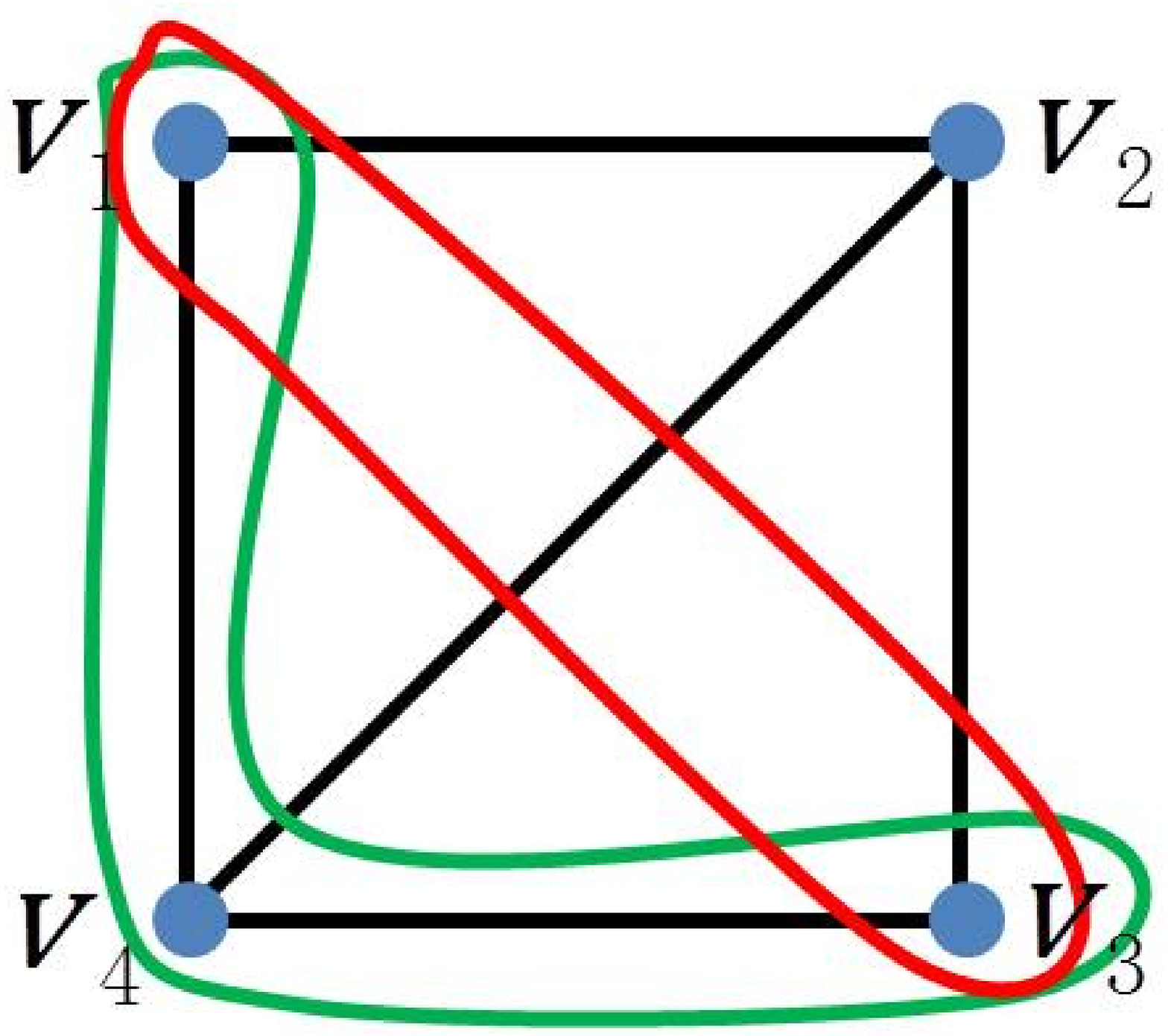}
\caption{\underline{General Networks}: If the link $(v_{1},v_{3})$ is removed from the original complete graph, we cannot get the measurement $[1\mbox{ }0\mbox{ }1\mbox{ }0]\bd_\cV$ any longer.}
\label{fig:node_not_complete}
\end{minipage}%
\hspace{0.1in}
\begin{minipage}[t]{0.48\linewidth}
\centering
\includegraphics[width=0.45\linewidth]{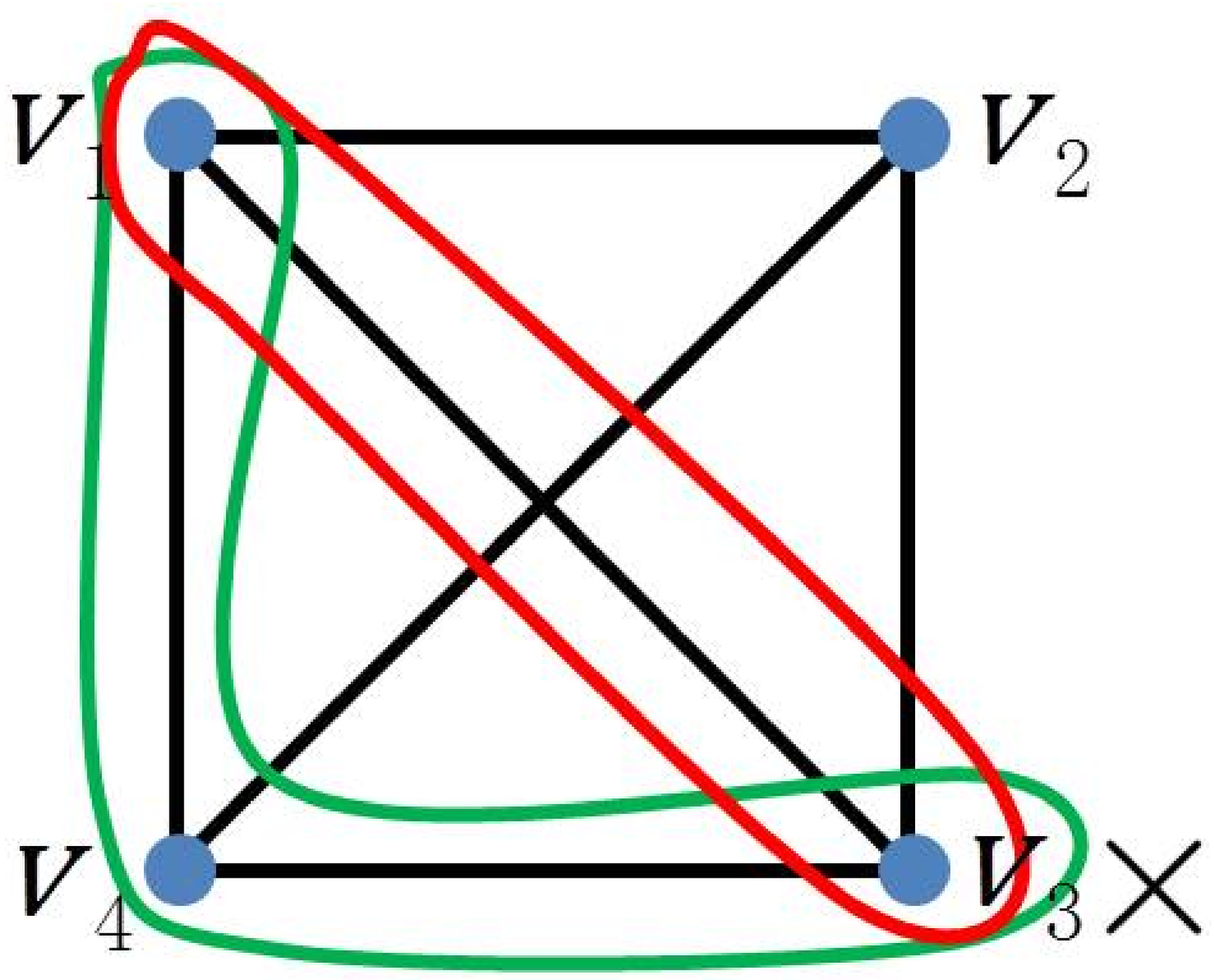}
\caption{\underline{Inaccessible Nodes}: If there is some constraint that we can not access to $v_{3}$ and $v_{3}$ is the destination of the paths for us to get the measurement, then any measurement in Fig. \ref{fig:nodedelay} is not available.}
\label{fig:node_not_access}
\end{minipage}
\end{figure}

In this section, we present some key observations and challenges that this paper focuses on. We begin with the observation that there is a high-level connection between the compressive sensing and the network tomography problem. As noted in the previous section,  network tomography can be treated as a problem of solving a system of linear equations. Under the assumption that the underlying unknown vector is sparse, it is natural to think of it as a compressive sensing problem~\cite{CandesT:2006}\cite{CanRT:06}\cite{Don:06}. Building on this intuition, network tomography can be formulated as the following compressive sensing problem: i) design a matrix $\bA$, ii) obtain delay measurements $\by=\bA\bd_\cV$ iii) reconstruct  $\bd_\cV$ from $\by$. Fig.~\ref{fig:nodedelay} illustrates this connection in a complete graph. Since each subset of nodes in a complete graph induces a connected subgraph, we can freely choose the locations of non-zero entries in each row of $\bA$. Then, any compressive sensing algorithm with $0$-$1$ measurement matrix {\cite{BerindeGIKS:2008} \cite{WeiyuH:2007}} can be applied to recover the vector $\bd_\cV$.
However, when we go beyond complete graphs and node measurements, it is not straightforward to apply compressive sensing directly. The network topology may impose constraints on implementable measurement matrices (See Figs.~\ref{fig:node_not_complete},\ref{fig:node_not_access},\ref{fig:link_complete}).

Xu {\it et al}.~\cite{XuMT:11} get around some of these problems by using random walks. One drawback of their approach is that it involves a factor of mixing time $T_\cN$ for both the number of measurements and path length. For networks without sufficient connectivity, mixing time may be very large, {\it e.g.}, cycle graph, $T_\cN={\cal O}(|\cE|^{2})$. In the following, we propose two news ideas that enable us to circumvent the above problem.

\begin{figure}[ht]
\centering
\includegraphics[width=0.5\linewidth]{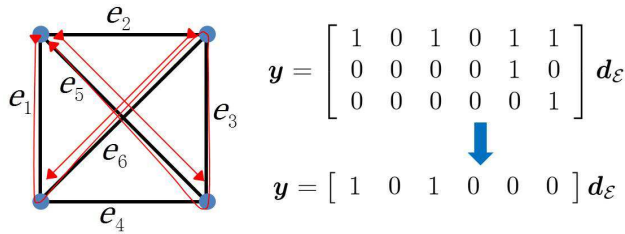}
\caption{\underline{Cancellation}: There are three paths in this graph:$\{e_{1}e_{6}e_{3}e_{5}\}$, $\{e_{5}\}$ and $\{e_{6}\}$. Triangles indicate the source and destination of a path. Correspondingly, we can derive three measurements $[1\mbox{ }0\mbox{ }1\mbox{ }0\mbox{ }1\mbox{ }1]\bd_\cE$, $[0\mbox{ }0\mbox{ }0\mbox{ }0\mbox{ }1\mbox{ }0]\bd_\cE$, and  $[0\mbox{ }0\mbox{ }0\mbox{ }0\mbox{ }1]\bd_\cE$. Subtracting the second and the third measurements from the first measurement, we get the measurement $[1\mbox{ }0\mbox{ }1\mbox{ }0\mbox{ }0\mbox{ }0]\bd_\cE$ which cannot be got by just one path.}
\label{fig:linkcancel}
\end{figure}

\begin{figure}[t]
\begin{minipage}[t]{0.49\linewidth}
\centering
\includegraphics[width=0.4\linewidth]{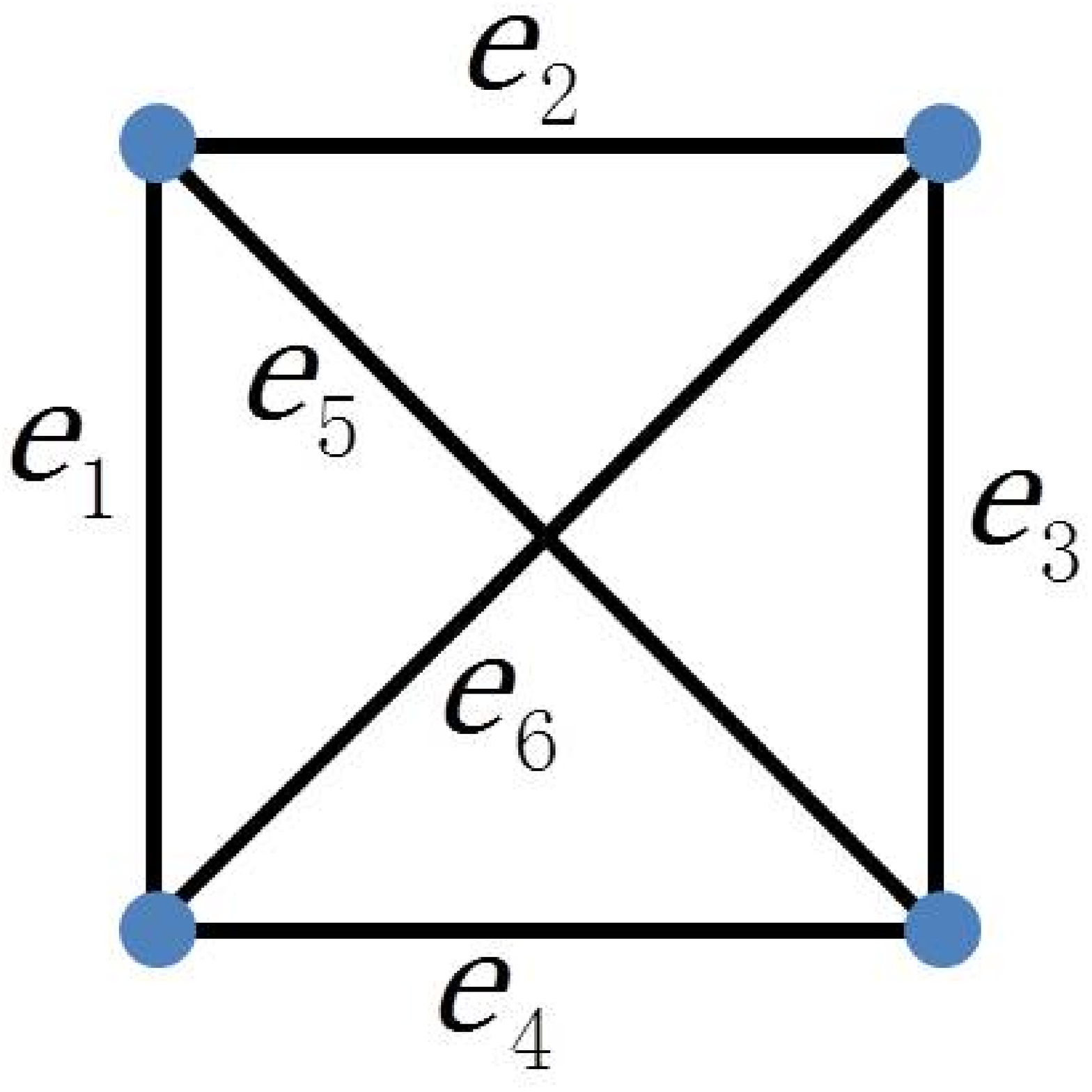}
\caption{\underline{Edge Delay Estimation}: We know that we can not get arbitrary measurement by one path even if the graph is complete. ({\it e.g.}, the measurement $[0\mbox{ }0\mbox{ }0\mbox{ }0\mbox{ }1\mbox{ }1]\bd_\cE$ cannot be got since there is no path just visiting $e_{5}$ and $e_{6}$.)}
\label{fig:link_complete}
\end{minipage}%
\hspace{0.1in}
\begin{minipage}[t]{0.49\linewidth}
\centering
\includegraphics[width=0.4\linewidth]{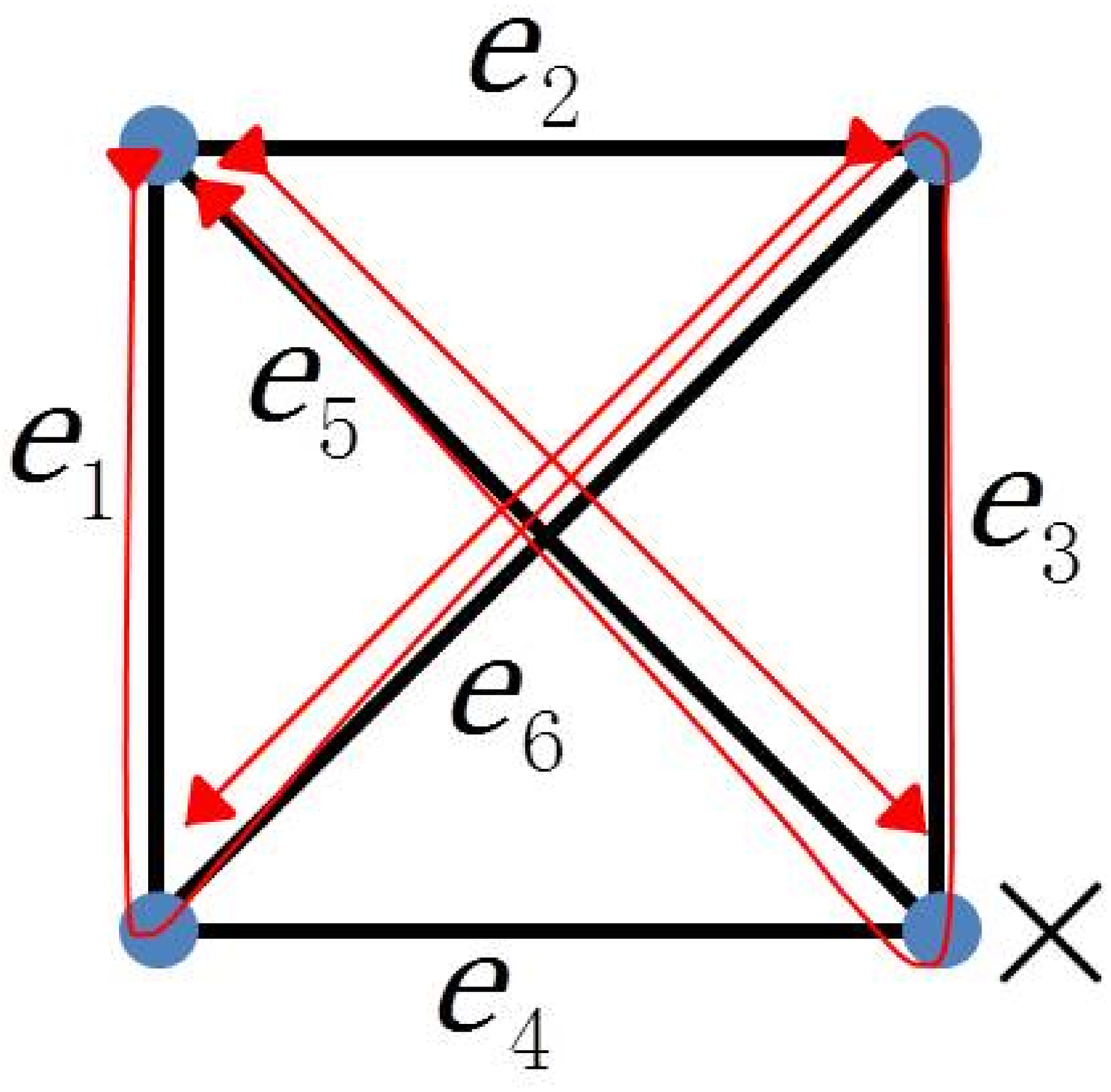}
\caption{\underline{Inaccessible Nodes}: The second measurement in Fig. \ref{fig:linkcancel} cannot be got since the $v_{3}$ which is the destination of the second path is not accessible (the same node identifier in Fig. \ref{fig:nodedelay}).}
\label{fig:link_not_access}
\end{minipage}
\end{figure}

\noindent \underline{\bf Idea 1}: {\em Cancellation enables selecting disconnected subsets of links and nodes.}
The idea here is similar to that used in \cite{MengWMA:2012} where they use the structure called hub to get arbitrary measurement matrix. However, they only consider the node delay model and special graphs which have $r$-partition. In this paper, we expand this approach to both link delay and node delay models. By considering correlated measurements, we can cancel out the contribution of the undesired links and nodes in a given measurement. Using this approach, we can mimic arbitrary measurements on general graphs. See Fig. \ref{fig:linkcancel} as an illustration. One drawback of the cancellation based approach is that if the selected measurement has too many disjoint components, then the number of measurements required is very large. {In Fig.~\ref{fig:linkcancel}, the number of cancellations is $2$.}

\noindent \underline{\bf Idea 2}: {\em Weighted measurements reduce the number of cancellations required and allow us to implement arbitrary integer valued matrices.}
\begin{figure}[bp]
\centering
\includegraphics[width=0.5\linewidth]{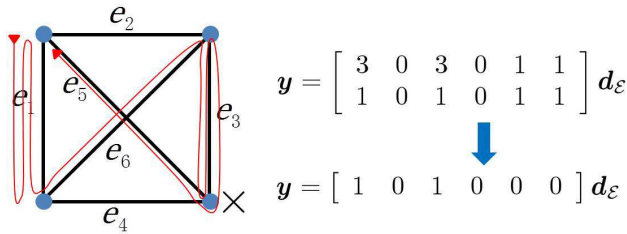}
\caption{\underline{Cancellation using weighted measurements}: To get the measurement $[1\mbox{ }0\mbox{ }1\mbox{ }0\mbox{ }0\mbox{ }0]\bd_\cE$, we design the paths as follows. First, we just follow the path $\{e_{1}e_{6}e_{3}e_{5}\}$, we get the measurement $[1\mbox{ }0\mbox{ }1\mbox{ }0\mbox{ }1\mbox{ }1]\bd_\cE$. Second, when visiting $e_{1}$ and $e_{3}$ for the first time, the probe does one more local loop for both links to get the measurement $[3\mbox{ }0\mbox{ }3\mbox{ }0\mbox{ }1\mbox{ }1]\bd_\cE$. Finally, we take the difference of these two measurements and divide the result by $2$. {Note that  1) Only one cancellation is required. 2) Even if $v_{3}$ is inaccessible, we still can achieve the two target measurements. 3) One additional local loops at $e_{1}$ in the second step (so that $e_{1}$ is visited $5$ times), allows us the measurement $[2\mbox{ }0\mbox{ }1\mbox{ }0\mbox{ }0\mbox{ }0]\bd_\cE$. Thus, controlling the number of local loops allows us to implement other ensembles of measurement matrices.}
}
\label{fig:loopycancel}
\end{figure}
The insight here is that if we have two paths along the same set of links, we can assign different weights for each link (or node) on these paths by performing local loops. Specifically, for a given set of weights on a subset of links (or nodes), we construct two measurements - a spanning measurement, and a weighted measurement. The spanning measurement is constructed by finding any path that visits through all the links (or nodes) in the desired subset at least once. The weighted measurement, then follows the same set of edges as the spanning measurement, but visits each link (or node) an additional number of times in accordance with the desired weight for that link (or node). Finally, we subtract the end-to-end delay for the weighted path from that of the spanning path to get an output that is proportional to the output of the corresponding compressive sensing problem (See Fig. \ref{fig:loopycancel}). These ideas enable us to reduce the network tomography problem to a compressive sensing problem with integer valued matrices. In the Section~\ref{sec:shofa}, we present an efficient compressive sensing algorithm with integer entries.

\section{Main Results}
In this section, we state the main results of this paper. Let $\rho\in\Omega(1)\cap o(|\cE|/\spar)$ be a design parameter.
\begin{theorem}[Compressive sensing via integer matrices] \label{thm:shofa}Let $M\in\mathbb{Z}^+$. There exists a constant $\cnst$ such that whenever $\nm>\cnst\spar\lceil\log{\samp}/\log{M}\rceil$, the ensemble of  $\mathbb{Z}_M$-valued  matrices $\{\bA_{\nm\times\samp}\}$ designed in Section~\ref{sec:shofa} and the \shofa reconstruction algorithm has the following properties:
\begin{enumerate}
\item Given $(\bA_{\nm\times\samp},\bA_{\nm\times\samp}\bd)$ as input, where $\bd$ is an arbitrary $\spar$-sparse vector in $\mathbb{R}^\samp$, $\shofa$ outputs a vector $\hat{\bd}\in\mathbb{R}^{\samp}$  that equals $\bd$ with probability at least $1-\cO(1/\spar)$ under the  distribution of $\bA_{\nm\times\samp}$ over the  ensemble $\{\bA_{\nm\times\samp}\}$.
\item Given $\bA_{\nm\times\samp}\bd$,  $\hat{d}$ is reconstructed in $\cO(\spar\lceil\log{n}/\log{M}\rceil)$ arithmetic operations.
\item Each row of $\bA_{\nm\times\samp}$ has $\cO(\samp/\spar)$ non-zeros in expectation.
\end{enumerate}
\end{theorem}
\hide{\begin{theorem}[Compressive sensing via integer matrices] \label{thm:shofa}Let $M\in\mathbb{Z}^+$. For some $\nm=\cO(\spar\lceil\log{\samp}/\log{M}\rceil)$, the ensemble of  $\mathbb{Z}_M$-valued  matrices $\{\bA_{\nm\times\samp}\}$ designed in Section~\ref{sec:shofa} and the \shofa reconstruction algorithm has the following properties:
\begin{enumerate}[leftmargin=.4em,itemindent=1.0em]
\item Given $(\bA_{\nm\times\samp},\bA_{\nm\times\samp}\bd)$ as input, where $\bd$ is an arbitrary $\spar$-sparse vector in $\mathbb{R}^\samp$, $\shofa$ outputs a vector $\hat{\bd}\in\mathbb{R}^{\samp}$  that equals $\bd$ with probability at least $1-\cO(1/\spar)$. Here, the error probability is derived from the probability distribution of $\bA_{\nm\times\samp}$ over the  ensemble $\{\bA_{\nm\times\samp}\}$.
\item {\textsf{\textsc{sho-fa-int}}} requires $\cO(\spar\lceil\log{n}/\log{M}\rceil)$ arithmetic operations.
\item The expected number of non-zero entries in each row of $\bA_{\nm\times\samp}$ is $\cO(\samp/\spar)$.
\end{enumerate}
\end{theorem}}
\begin{theorem}[Network tomography for link congestion]~\label{thm:edge} Let $\cN=(\cV,\cE)$ be an undirected network of diameter $D$ such that at most $\spar$ have unknown {non-zero} link delays. Let $M\in\mathbb{Z}^+$ Then, the \fr algorithm has the following properties:
\begin{enumerate}

\item \fr requires $\cO(\rho\spar\lceil\log{|\cE|}/\log{M}\rceil)$ measurements.
\item For every edge delay vector $\bd_\cE\in\mathbb{R}^{|\cE|}$,\fr outputs $\hbd_\cE$ that equals $\bd_\cE$ with probability $1-\cO(1/\rho\spar)$.
\item The \fr reconstruction algorithm requires $\cO(\rho\spar\lceil\log{|\cE|}/\log{M}\rceil)$ arithmetic operations.
\item The number of links of $\cN$ traversed by each test measurement packet in \fr is $\cO(D|\cE|/\rho\spar)$ and the total number of hops for each packet is $\cO(DM|\cE|/\rho\spar)$.
\end{enumerate}
 \end{theorem}
\hide{\begin{theorem}[Network tomography for link congestion]~\label{thm:edge} Let $\cN=(\cV,\cE)$ be an undirected network of diameter $D$ such that at most $\spar$ have unknown {non-zero} link delays. Let $M\in\mathbb{Z}^+$ Then, the \fr algorithm described above has the following properties:
\begin{enumerate}

\item \fr requires $\cO(\spar\lceil\log{|\cE|}/\log{M}\rceil)$ measurements.
\item For every edge delay vector $\bd_\cE\in\mathbb{R}^{|\cE|}$,\fr outputs $\hbd_\cE$ that equals $\bd_\cE$ with probability $1-\cO(1/\spar)$.
\item The \fr reconstruction algorithm requires $\cO(\spar\lceil\log{|\cE|}/\log{M}\rceil)$ arithmetic operations.
\item The number of links of $\cN$ traversed by each test measurement packet in \fr is $\cO(D|\cE|/\spar)$ and the total number of hops for each packet is $\cO(DM|\cE|/\spar)$.
\end{enumerate}
 \end{theorem}}


\begin{definition}[Isolated congested node] A congested node is called isolated if there exists one of its neighbours which is not congested.
\end{definition}

\begin{theorem}[Network tomography for node congestion] \label{thm:node}Let $\cN=(\cV,\cE)$ be an undirected network of diameter $D$ such that at most $\spar$ have unknown {non-zero} node delays and all congested nodes are isolated. Let $M\in\mathbb{Z}^+$ Then, the \fr algorithm has the following properties:
\begin{enumerate}

\item \fr requires $\cO(\rho\spar\lceil\log{|\cV|}/\log{M}\rceil)$ measurements.
\item For every edge delay vector  { $\bd_{\cV}\in\mathbb{R}^{|\cV|}$}, \fr outputs $\hbd_\cV$ that equals $\bd_\cV$ with probability $1-\cO(1/\rho\spar)$.
\item The \fr reconstruction algorithm requires $\cO(\rho\spar\lceil\log{|\cV|}/\log{M}\rceil)$ arithmetic operations.
\item The number of links of $\cN$ traversed by each test measurement packet in \fr is $\cO(D|\cV|/\rho\spar)$ and the total number of hops for each packet is $\cO(DM|\cV|/\rho\spar)$.
\end{enumerate}
\end{theorem}
\section{\shofa algorithm for Compressive Sensing}\label{sec:shofa}
We begin by describing a new compressive sensing algorithm \shofa which has several properties that are desirable for our application. \shofa is related to the {\textsf{\textsc{sho-fa}}} algorithm -- originally developed in the unconstrained compressive  sensing setting~\cite{BakJCC:12} -- but differs from it in that the non-zero entries of the sensing matrix $\bA$ are constrained to be positive integers less than or equal to some $M\in\bbn$. \footnote{Reference \cite{BakJCC:12} proposes a design of matrix $\bA_{\meas\times\samp}$ such that given $\by=\bA\bd$ for a $\spar$-sparse vector $\bd\in\bbr^\samp$, a reconstruction $\hat{\bd}$ can be obtained in $\cO(\spar)$ steps by using a measurement vector of length $\meas=\cO(\spar)$. A key requirement of this design is that both the locations of non-zero entries of $\bA$ as well as their values may be arbitrarily chosen. In particular, the non-zero entries of $\bA$ are chosen to be unit norm complex numbers.}

Let $\{\graph_{\samp,\meas}\}_{\samp,\meas\in\bbn}$ be an ensemble of left-regular bipartite graphs, where  each $\graph_{\samp,\meas}$ is a bipartite graph with left vertex set $\set{1,2,\ldots,\samp}$ and right vertex set $\set{1,2,\ldots,\meas}$. For each left vertex $\xj\in\{1,2,\ldots,\samp\}$, we pick three distinct vertices uniformly at random from the set of right vertices $\{1,2,\ldots,\meas\}$.


%
%
%
%
\noindent{\em Measurement Design:}\label{sec:design}
Let $\zeta(\cdot)$ be the Riemann zeta function. Let $\eL\in\bbn^+$ such that $M^\eL/\zeta(\eL)\geq 3\samp$ and let $[M]$ denote the set $\set{1,2,\ldots,M}$. Given the graph $\graph_{\samp,\meas}$, we design a $\eL\meas\times\samp$ measurement matrix $\bA(=\bA_{\eL\meas\times\samp})$ as follows. First, we partition the rows of $\bA$ into $\meas$ groups of rows, each consisting of $\eL$ consecutive rows as follows. {$$\bA= \left[\begin{array}{c:c:c:c}
a^{(1)}_{11}&a^{(1)}_{12}&\dots & a^{(1)}_{1\samp}\\
\vdots &\vdots &\ddots&\vdots\\
a^{(\eL)}_{11}&a^{(\eL)}_{12}&\dots & a^{(\eL)}_{2\samp}\\
\hdashline
a^{(1)}_{21}&a^{(1)}_{22}&\dots & a^{(1)}_{2\samp}\\
\vdots &\vdots &\ddots&\vdots\\
a^{(\eL)}_{21}&a^{(\eL)}_{22}&\dots & a^{(\eL)}_{2\samp}\\
\hdashline
\vdots &\vdots &\ddots&\vdots
\end{array}\right]$$ }Let $a_{\yi \xj}^{(\el)}$ be the $\el$-th row entry in the $\xj$-th column of the $\yi$-th group and let $\pmb{a}_{\yi \xj}=[a_{\yi \xj}^{(1)}a_{\yi \xj}^{(2)}\ldots a_{\yi \xj}^{(\eL)}]^T$.  First, for each $(\yi,\xj)$ not in $\graph_{\samp,\meas}$, we set $\pmb{a}_{\yi \xj}=0^{R}$. Next, we randomly chose $3\samp$ distinct values from the set $\mathcal{C}\triangleq\left\{[c_1,c_2,\ldots,c_\eL]^{T}\in(\mathbb{Z}_M)^\eL:\mbox{gcd}(c_1, c_2, \ldots, c_\eL)=1\right\}$ and use these to set the values of $\pmb{a}_{\yi \xj}$ for each edge $(\yi,\xj)$ in $\graph_{\samp,
\meas}$.  {The assumption that $M^\eL/\zeta(\eL)\geq 3\samp$ ensures that such a sampling is possible. }{We skip the proof of Lemma~\ref{lem:coprimes} here and refer the reader to~\cite{BakJCC:12} for the proof.

\begin{lemma}[]~\cite[Lemma~6]{BakJCC:12}\label{lem:coprimes} For $M$ large enough, $M^\eL/\zeta(\eL)\leq |\mathcal{C}|\leq M^\eL$. \end{lemma}
}


{The output of the measurement is a $\eL\meas$-length vector $\by=\bA\bd$. Again, we partition $\by$ into $\meas$ groups of $\eL$ consecutive rows each, and denote the $\yi$-th sub-vector as $\by_\yi$. Note each $\by_\yi\in\bbr^\eL$ follows the relation $\by_\yi=[a_{\yi 1}a_{\yi 2}\ldots a_{\yi \samp}]\bd$.}

\noindent{\em Reconstruction algorithm:}
The decoding process is essentially equivalent to the {}``peeling
process'' to find $2$-core in uniform hypergraph~\cite{Bloom,Molloy:2005}.  The decoding takes place over $\cO(\spar)$ iterations. \hide{Let $S(\bd)=\set{\xj\in \{1,2,\ldots\samp\}: d_\xj\neq 0}$.} The decoding algorithm is very similar to~\cite[Section IV-B]{BakJCC:12}. In each iteration, we find one non-zero undecoded $d_\xj$ with a constant probability after locating a right node that is connected to exactly one non-zero left node. After decoding the non-zero $d_\xj$ for the current iteration, we cancel out the contribution of $d_\xj$ from all measurements and proceed iteratively.
{
To describe the peeling process, we define {leaf} nodes as follows.

\begin{definition}[$S$-leaf node] A right node $\yi$ is a leaf node for $S$ if $\yi$ is connected to exactly one $\xj\in S$ in the graph $\graph_{\samp,\meas}$.
\end{definition}
First, the algorithm initializes the {\em reconstruction vector} $\hbd(1)$ to the all zeros vector $0^{\samp}$, the {\em residual measurement vector} $\tilde{\by}(1)$ to $\by$, and the {\em neighbourly set} $\cD(1)$ to be the set of all right nodes for which $y_i$ does not equal $0^\eL$. In each iteration $\iter$, the decoder picks a right node $\yi(\iter)$ from the current neighbourly set $\cD(\iter)$ and checks if only one left node contributes to the value of $\left(\tilde{y}(\iter)\right)_{\yi(\iter)}$. If so, it identifies $\yi(\iter)$ as a {\em leaf} node, decodes delay value at the corresponding parent node, and updates $\cD(\iter+1),\tilde{\by}(\iter+1)$, and $\hbd(\iter+1)$ for the next iteration. The decoder terminates when the residual measurement vector becomes zero. See~\cite[Section IV-B]{BakJCC:12} for a detailed description.}
%
%
%
%
%
%
%
%
%

%
%
%
%

{
Next, we prove the performance guarantees of \shofa as claimed in Theorem~\ref{thm:shofa}. Let $\spar=\spar(\samp)$ grow as a function of $\samp$. We show that the algorithm presented above correctly reconstructs the vector $\hbd$ with a high probability over the ensemble of matrices $\{\bA_{\eL\meas\times\samp}\}$. To this end, we first note  that if $\meas=\Omega(\spar)$, the ensemble of graphs $\{\graph_{\samp,\meas}\}$ satisfies the following ``many leaf nodes'' as shown in the following lemma.
\begin{lemma}[Many leaf nodes]~\cite[Lemma~2]{BakJCC:12}\label{lem:leaf} Let $S$ be a subset of the left nodes of the $\graph_{\samp,\meas}$ and let $N(S')$ be the set of right neighbours of a set $S'$. If $|S|\leq\spar$ then with probability $1-\cO(1/\spar)$, for every $S'\subseteq S$, $N(S')$ contains at least $|N(S')|/2$ $S'$-leaf nodes.
\end{lemma}

\proofof{Theorem~\ref{thm:shofa}}
Let $S(\bd)=\{\xj\in\{1,2,\ldots,\samp\}:d_\xj\neq 0\}$. By Lemma~\ref{lem:leaf}, with probability $(1-\cO(1/\spar))$, all its subsets $S'$ of $S(\bd)$ have at least twice as many leaf neighbours as the the number of elements in the $S'$. Therefore, in each iteration, the probability of picking a leaf node is at least half. Next, we note that in each iteration that we pick a leaf node, the probability of identifying as one and finding it's left neighbour correctly is $1$. This is true because the weight vectors $\ba_{\yi\xj}$'s corresponding  different neighbours of a given right node $\yi$ are different and for a leaf node $\yi$ with the sole non-zero neighbour $\xj$, the output value $\by_\yi$ exactly equals to $d_\xj\ba_{\yi\xj}$.

Next, we argue that if $\yi$ is not a leaf node, then the probability of it being declared a leaf node in any iteration is $\cO(1/\samp)$. Note that for this error event to occur for a right node $\yi$, it must be true that  $\sum_{\xj'\in N(\yi)}d_{e'}a_{\yi\xj'}=d'' a_{\yi\xj''}$ for some $d''\in\bbr$ and $\xj''$ connected to $\yi$. Since all the measurement weights  are chosen randomly, by Schwartz-Zippel lemma~\cite{Sch:80,Zip:79}, the probability of this event is $\cO(1/\samp)$, which is $o(1/\spar)$.


Since the probability of picking a leaf node at any iteration is at least $1/2$, the expected number of iterations before a new leaf node is picked is upper bounded by $2$.  Since there are at most $\spar$ non-zero $d_\xj$'s, in expectation, the algorithm terminates in $\cO(\spar)$ steps. Further, since the probability of finding a leaf in each iteration is independent of other iterations, by applying standard concentration arguments, the total number of iterations required is upper bounded by $2\spar$ in probability.

Finally, to compute the decoding complexity, note that each iteration requires a constant number arithmetic operations over vectors in $[M]^\eL$, which in turn can be decomposed into $\cO(\eL)$ arithmetic operations over integers. Therefore, the total number of integer operations required is $\cO(\eL\spar)=\cO(\spar\lceil\log{\samp}/\log{M}\rceil)$. Finally, we note that since each left node in $\graph_{\samp,\meas}$ has exactly $3$ right neighbours which are picked uniformly among all right nodes and independently across different left nodes, with a high probability, each right node has no more than $4\samp/\meas$ left neighbours. This can be proved by first computing the expected number of left neighbours for a right node and then applying Chernoff bound to concentrate it to close to its expectation. This shows that, with a high probability, the number of non-zero entries in each row of $\bA$ is $\cO(\samp/\spar)$.
\endproof}
\section{The \fr algorithm} \label{sec:frantic}
\noindent {\em\phantom{a}A. Link Delay Estimation:}\label{sec:linkdelay}
We define a {\em path} $\cP$ of length $T$ over the network $\cN=(\cV,\cE)$ as a sequence $(e_1,e_2,\ldots,e_{T})=\left((v_1,v_2),(v_2,v_3),\ldots (v_{T},v_{T+1})\right)$ such that $e_t\in\cE$ for $t=1,2,\ldots,T$. For a given path $\cP$, we define the multiplicity $W(\cP,e)$ of a link $e\in \cE$ as the number of times $\cP$ visits $e$. Let $\Delta(\cP)$ be the end-to-end delay for a path $\cP$.


\begin{definition}[$\pmb{w}$-spanning measurement]Given a measurement weight vector $\pmb{w}=[w_1 w_2 \ldots w_{|\cE|}]$, and  a {$\pmb{w}$-\em{spanning measurement}} is a path $\cP=(e_1, e_2.,\ldots e_T)$ in the network $\cN$ such that $\cP$ visits each $e$ in $\{e:w_e\neq 0\}$ at least once.
\end{definition}

\begin{definition}[$(\pmb{w},\cP)$-weighted measurement]Given a measurement weight vector $\pmb{w}$ and a $\pmb{w}$-spanning measurement $\cP=(e_1,e_2,\ldots e_T)$, a {\em $(\pmb{w},\cP)$-weighted measurement} is a path $\pmb{Q}=(e_1', e_2', \ldots e_H')$ in the network $\cN$ such that $W(\pmb{Q},e)=W(\cP,e)+2w_e$ for each link $e$.
\end{definition}

Observe that the end-to-end delay for a $\pmb{w}$-spanning measurement $\cP$ is equal to $\Delta(\cP)=\sum_{e\in\cE}W(\cP,e)d_e$, and that for a $(\pmb{w},\cP)$-weighted measurement is equal to
\begin{equation}\label{eq:weighted}
\Delta(\pmb{Q})=\sum_{e\in\cE}W(\pmb{Q},e)d_e\hide{=\sum_{e\in\cE}W(\cP,e)d_e + 2\sum_{e\in\cE}w_ed_e\nonumber\\
&=&}= \Delta(\cP)+2\sum_{e\in\cE}w_ed_e.
\end{equation}
{\em Proof of Theorem~\ref{thm:edge}:}
To prove Theorem~\ref{thm:edge}, we start with a measurement matrix $\bA$ drawn according to the \shofa construction for Theorem~\ref{thm:shofa}. For each row of the measurement matrix, we construct two paths in the network - a spanning measurement and a weighted measurement. Next, we subtract the end-to-end delay for the spanning measurement from the weighted measurement to get an output that is exactly twice the measurement output corresponding to the compressive sensing measurement using measurement matrix $\bA$. Thus, we can apply the \shofa reconstruction algorithm from Section~\ref{sec:shofa} to reconstruct the delay vector $\bd_{\cE}$. More precisely, Let $\bA$ be a $\eL\meas\times\samp$ matrix drawn from the ensemble of Section~\ref{sec:shofa}, where $\eL=\cO(\lceil\log{\samp}/\log{M}\rceil)$ and $n=|\cE|$.\\
\noindent{\em \underline{Measurement Design:}} Let $\ba(\yi)=[a_{\yi 1}a_{\yi 2}\ldots a_{\yi \samp}]$ be the $\yi$-th row of $\bA$. Consider network measurements $\cP(\yi)$ and $\cQ(\yi)$ defined as follows. Let $\cP(\yi)$ be an {\em $\ba(\yi)$-spanning measurement }obtained by picking the links in $\{e:\ba(\yi)\neq 0\}$ one-by-one and finding a path from one link to another. By the definition of the diameter of the graph, there exists a path of length at most $D$ between any pair of links. Therefore, there exists a path $\cP(\yi)=\left((v_1,v_2),(v_2,v_3),\ldots,(v_T,v_{T+1})\right)$ of length $T=\cO(D\samp/\spar)$ that covers all the $\cO(\samp/\spar)$ vertices that have non-zero components in $\ba(\yi)$.

\begin{SCfigure}[2.6]
\centering
\captionsetup{width=0.76\linewidth}
\caption{\underline{Isolated Node}: For a subgraph with $5$ vertices and $4$ links, all vertices are congested. $v_{1}$ is not isolated since all of its neighbors are congested. Suppose there is a local loop involving $v_1$, $v_{2}$, and $e_1$. For link measurement, only the delay of $e_1$ is added to the weighted measurement. However, for the node measurement, the delays of $v_1$ and $v_2$ are both added to the weighted measurement. The delay of $v_{2}$ will not be canceled by the corresponding spanning measurement.
}\includegraphics[width=0.25\linewidth]{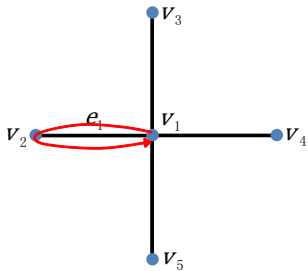}

\label{fig:isolate}
\end{SCfigure}

Next, let $\cQ(\yi)=\left(e_1',e_2'\ldots, e_{T'}'\right)$ be a {\em $(\cP(\yi),\ba(\yi))$-weighted measurement}  of length $T'=T+2\sum_{e\in\cE}a_e(\yi)$ as follows.
Let $e_1'=(v_1,v_2)$. If $a_{(v_1,v_2)}(i)\neq 0$, we traverse the edge $(v_1,v_2)$ an additional  $2a_{(v_1,v_2)}(\yi)$ times by going in the forward direction, {\em i.e.} on $(v_1,v_2)$, and the reverse direction, {\em i.e.} on $(v_2,v_1)$, an additional $a_{(v_1,v_2)}(\yi)$ times each. Thus, for $\tau=1,3,5,\ldots,2a_{(v_1,v_2)}(\yi)+1$, we set $e'_\tau=(v_1,v_2)$ and for $\tau=2,4,\ldots,2a_{(v_1,v_2)}(\yi)$, we set $e'_\tau=(v_2,v_1)$.  Next, if $v_3=v_1$, {\em i.e.,} we have already visited $e_2$, we traverse the link we traverse the link $e_2$  once more, else we traverse it $a_{(v_2,v_3)}(\yi)+1$ times in the forward direction and $a_{(v_2,v_3)}(\yi)$ times in the reverse direction, {\em i.e.},  for $\tau=2a_{(v_1,v_2)}(\yi)+2,2a_{(v_1,v_2)}(\yi)+4,\ldots,2a_{(v_1,v_2)}(\yi)+2a_{(v_2,v_3)}(\yi)+2$, we set $e_\tau'=(v_2,v_3)$ and for $\tau=2a_{(v_1,v_2)}(\yi)+3,2a_{(v_1,v_2)}(\yi)+5,\ldots,2a_{(v_1,v_2)}(\yi)+2a_{(v_2,v_3)}(\yi)+1$, we set $e_\tau'=(v_3,v_2)$. We continue this process for each link $(v_t,v_{t+1})$ in the path $\cP(\yi)$, {\em, i.e.}, if $(v_t,v_{t+1})$ has been visited already in either the forward or reverse direction by $\cQ(\yi)$, we add it to $\cP(\yi)$ only once, else, we traverse it an additional $a_{(v_t,v_{t+1})}(\yi)$ times in each direction. Therefore, $\cQ(\yi)$ visits every edge $e\in\cE$ a total of $2a_e(\yi)$ times more than $\cP(\yi)$ does.\\
\noindent{\em \underline{Reconstructing $\bd_\cE$:} } Next, we measure the end-to-end delays for the paths $\cP(\yi)$ and $\cQ(\yi)$ for each $\yi=1,2,\ldots,\eL\meas$ and let $y_{\yi}=(\Delta(\cQ(\yi))-\Delta(\cP(\yi)))/2$. From equation~\eqref{eq:weighted}, it follows that $y_{\yi}=\sum_{e\in\cE}a_{\yi e}d_e$. Note that this exactly equals the output of a compressive sensing measurement with $\bd$ as the input vector, $\bA$ as the measurement matrix, and $\by$ and the measurement output vector. Using this observation, we input the vector $\by$ to the \shofa algorithm \hide{of Section~\ref{sec:shofa} }to correctly reconstruct $\bd$ with probability $1-\cO(1/\spar)$. The guarantees on the decoding complexity follow from the decoding complexity of the \shofa algorithm \hide{as the only additional step is subtracting $\Delta(\cP(\yi))$ from $\Delta(\cQ(\yi))$ for each $\yi=1,2,\ldots,\eL\meas$, which is accomplished using $\cO(\eL\meas)$ arithmetic operations} and that on the total number of hops follows by noting that each link in a measurement path may be visited at most $2M$ times.\endproof
\noindent\phantom{a}{\em B. Node Delay Estimation:} The measurement design and the decoding algorithm for node delay estimation proceeds in a similar way to the link delay estimation algorithm of Section~\ref{sec:linkdelay}. The difference here is that instead of assigning weights to links in a path, our design assigns weights to nodes in a path by visiting each node repeatedly. We skip the proof of Theorem~\ref{thm:node} here as it essentially follows from the technique used in the proof of Theorem~\ref{thm:edge}.
The only difference is that for node delay estimation we add the isolation assumption. If there exists one congested node, $v\in\cV$, whose neighbors are all congested nodes, then we are not able to generate the measurement involving $v$ by subtracting the weighted measurement from the spanning measurement. The reason is that each local loop involving $v$ adds one more delay corresponding to one of its congested neighbor. However, this problem doesn't happen in the edge delay measurements. (See Fig. \ref{fig:isolate})

\section{Exploiting network structure}
\subsection{Reducing Path Lengths through Steiner Trees: }
\label{sec:steiner}
 One drawback of the approaches presented in the previous section is that even though on an average, each row of $\bA$ contains only $\cO(\samp/\rho\spar)$ non-zero entries, our upper bound on the path length relies on worst case pairwise paths for each pair of successive edges to be measured. In this Section, we propose a Steiner Tree based approach to design the measurement paths given a measurement matrix $\bA$.


\begin{definition}[Steiner Tree] Let $\cS\subseteq \cV$. We say that $\cT\subseteq \cE$ is a Steiner Tree for $\cS$ if $\cT$  has the least number of edges among all subsets of $\cE$ that form a connected graph that is incident on every $v\in\cS$. Let $L(\cS)$ be the length of a Steiner Tree for $\cS$.
\end{definition}

For every $s\in\mathbb{Z}^+$, let $$L^*(s)\triangleq \max_{\mathclap{\substack{\cS:\cS\subseteq \cV\\|\cS|\leq s}}}\;L(\cS).$$

Note that, in general, $L^*(s)\leq Ds$. Further, in many graphs of practical interest, $L^*(s)\ll Ds$. For example, in a line graph with $\samp$ vertices, $L^*(s)$ is at most $\samp$, while $Ds$ maybe as large as $\cO(\samp s)$. Using this observation, we may further improve the performance guarantee of our algorithm. We note that it suffices to find a Steiner Tree that passes through all links specified by a given row of the measurement matrix $\bA$. Also, we already know that, with a high probability, the number of non-zero entries in each row of $\bA$ is $\cO(\samp/\rho\spar)$. Thus, in general, the number of links traversed by each link (or node) delay measurement is $\cO(L^*(s))$ where $s={\cal O}(|\cE|/\rho\spar)$ (or ${\cal O}(|\cV|/\rho\spar)$ respectively) is the number of non-zero entries in the measurement. This proves the following assertion.
\begin{theorem}[Network tomography for link/node congestion using Steiner Trees]\label{thm:link_steiner} \hide{Let $\cN=(\cV,\cE)$ be an undirected network such that at most $\spar$ links have unknown {non-zero} link delays. Let $M\in\mathbb{Z}^+$ Then, the number of measurements, probability of correct reconstruction, and number of arithmetic operations required for the \fr algorithm described in Section~\ref{sec:steiner}  have the same guarantees as in Theorem~\ref{thm:edge} and~\ref{thm:node}. Further, the number of links of $\cN$ traversed by each measurement is at most $\cO(L^*(s))$ where $s={\cal O}(|\cE|/\rho\spar)$ is the number of non-zero entries in the measurement and the total number of hops is $\cO(ML^*(s))$.} For the setting of Theorem~\ref{thm:edge}, the number of links of $\cN$ traversed by each measurement of \fr is at most $\cO(L^*(s))$ where $s={\cal O}(|\cE|/\rho\spar)$ is the number of non-zero entries in the measurement and the total number of hops for each measurement is $\cO(ML^*(s))$.\end{theorem}
\hide{
\begin{theorem}[Network tomography for node congestion using Steiner Trees]\label{thm:node_steiner} Let $\cN=(\cV,\cE)$ be an undirected network such that at most $\spar$ have unknown non-zero node delays and all congested nodes are isolated. Let $M\in\mathbb{Z}^+$ Then, the number of measurements, probability of correct reconstruction, and number of arithmetic operations required for the \fr algorithm described in Section~\ref{sec:steiner}  have the same guarantees as in Theorem~\ref{thm:node}. Further,  the number of links of $\cN$ traversed by each measurement is at most $\cO(L^*(s))$ where $s={\cal O}(|\cV|/\rho\spar)$ is the number of non-zero entries in the measurement and the total number of hops is $\cO(ML^*(s))$.\end{theorem}}

\noindent \underline{Remark:} There exist polynomial-time approximation schemes with a performance ratio decreased from $2$ to $1.55$ by a series of works~\cite{TakM:1980,Zel:1993,BerR:1994,Zel:1996,ProS:1997,KarZ:1997,HouP:1999,RobZ:2000}.

\subsection{Average length of Steiner Trees: }In Theorem~\ref{thm:link_steiner}, we analyzed the length of measurement paths in terms of the {worst-case} length of Steiner trees that contain an arbitrary subset of $s$ links (resp. nodes). However, on an average, however, this may be too conservative an estimate.


\begin{definition}[Average length of Steiner tree] For every $s\in\bbn^+$, let \[
\overline{L}(s)\triangleq\frac{\sum_{{\substack{\cS:\cS\subseteq\cV\\|\cS|=s}}}L({\cal S})}{|\{\cS\subseteq\cV:|\cS|=s\}|}
\] denote the average length of Steiner tree.
\end{definition}

\begin{figure}[bp]
\centering
\captionsetup{width=0.8\linewidth}\includegraphics[width=0.59\linewidth]{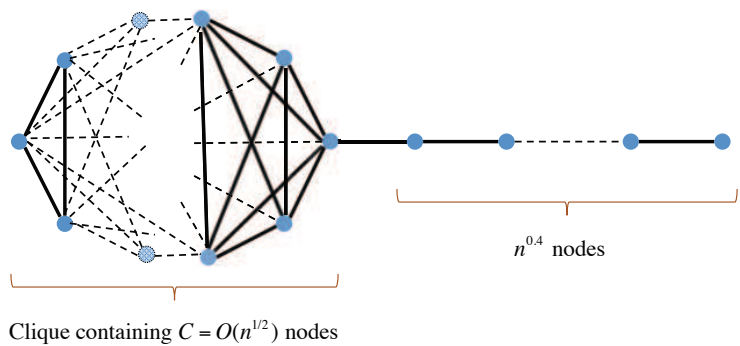}
\caption{\underline{Worst-case vs Average length of Steiner trees}: Consider a network of $n$ links. The network has two parts - a clique consisting of $C=(1+\sqrt{1+8(n-n^{0.4})})/2$ fully connected nodes and a line consisting of $n^{0.4}$ nodes. Let the number of congested links in the network be $k=n^{0.95}$. Thus, each measurement path has to cover a set of links of size $\cO(n^{0.05})$ specified by \shofa. In the worst case, such a set can include the two ends of the linear part. Thus, in the worst case, the length of the Steiner tree can exceed $n^{0.4}$. However, we note that the \shofa algorithm picks the measurement nodes uniformly at random. Thus, the probability of picking even one edge from the linear part of the network is ${\cal O}(n^{0.45}/n^0.5)={\cal O}(n^{-0.5})$ by the union bound. Therefore, on an average, the length of Steiner tree is at most $\cO(n^{-0.5}\times n^{0.4})=\cO(n^{0.35})$, which is lower than the worst-case length of a Steiner tree covering $\cO(n^{0.05})$ links in the network.}
\label{fig:skewednetwork}
\end{figure}

In the example shown in Fig.~\ref{fig:skewednetwork}, we argue that, with a high probability, the length of paths required is upper bounded by $\overline{L}(s)$ which may be significantly smaller than $L^*(s)$. \vspace{-1em}
\hide{\begin{definition}[Average length of Steiner tree] For every $s\in\bbn^+$, let \[
\overline{L}(s)\triangleq\frac{\sum_{{\substack{\cS:\cS\subseteq\cV\\|\cS|=s}}}L({\cal S})}{|\{\cS\subseteq\cV:|\cS|=s\}|}
\] denote the average length of Steiner tree.
\end{definition}}
\begin{spacing}{0.9}
\end{spacing}
\subsection{Network decomposition: }
\begin{figure}[ht!]
\centering
\includegraphics[width=0.4\linewidth]{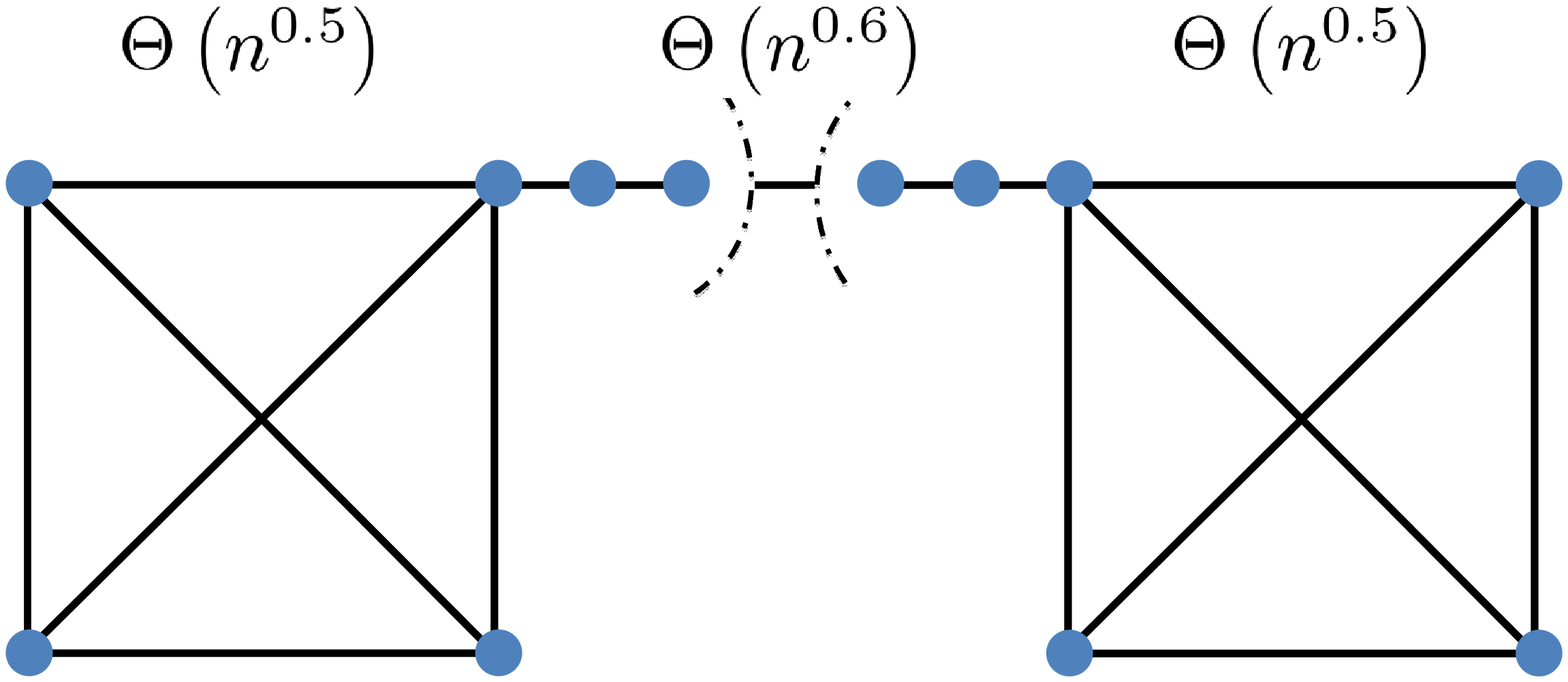}
\caption{\underline{Network Decomposition}: The network consists of three parts -- two complete graphs with $\Theta\left(n^{0.5}\right)$ vertices and a line graph with $\Theta\left(n^{0.6}\right)$ vertices. It follows that there are $\Theta(n)$ links in each of the two complete graphs and $\Theta\left(n^{0.6}\right)$ links in the line graph. For each link measurement, with high probability, two links involved locate in each of two complete graphs. Therefore, the average length of Steiner tree is at least $\Theta\left(n^{0.6}\right)$. If we decompose the original network into two subgraphs as shown in the figure and do the link delay estimation on them separately, the average length of Steiner tree becomes at most $\Theta\left(n^{0.5}\right)$ which is smaller than $\Theta\left(n^{0.6}\right)$.
}
\label{fig:decomposition}
\end{figure}

Since we already know the topology of the network, exploring the structure of the topology may help us to reduce the path length of each measurement. In Fig.~\ref{fig:decomposition}, we illustrate how to reduce the length of Steiner tree by network decomposition.

\section{Acknowledgements}
The work described in this paper was partially supported by a grant from University Grants Committee of the Hong Kong Special Administrative Region, China (Project No. AoE/E-02/08), a grant from the Microsoft-CUHK Joint Laboratory for Human-centric Computing and Interface Technologies, and an SHIAE grant.

\bibliographystyle{IEEEtran}
\bibliography{IEEEabrv}

\end{document}